\documentclass{PoS}
\usepackage[sort&compress,numbers]{natbib}
\usepackage{mathtools}
\usepackage{amsfonts} 
\usepackage{amssymb} 
\usepackage{amsmath} 
\usepackage{graphicx} 
\usepackage{latexsym} 
\usepackage{verbatim} 

\newcommand{\K}{{\mathcal K}}
\newcommand{\cM}{{\mathcal M}}
\newcommand{\CM}{{\rm CM}}
\newcommand{\df}{{\rm df}}
\newcommand{\ttil}{{\tilde 2}}

\newcommand{\HSBS}[0]{Hansen:2016ync}

\newcommand{\HSTH}[0]{Hansen:2016fzj}
\newcommand{\BHSQC}[0]{Briceno:2017tce}
\newcommand{\BHSnum}[0]{Briceno:2018mlh}
\newcommand{\HSQCa}[0]{Hansen:2014eka}
\newcommand{\HSQCb}[0]{Hansen:2015zga}
\newcommand{\AkakiBS}[0]{Meissner:2014dea}
\newcommand{\Akakia}[0]{Hammer:2017uqm}
\newcommand{\Akakib}[0]{Hammer:2017kms}

\newcommand{\Luscher}[0]{Luscher:1986pf,Luscher:1990ux}
\newcommand{\Mai}[0]{Mai:2017bge}

\title{Progress report on the relativistic three-particle quantization condition}

\ShortTitle{Progress on the relativistic three-particle quantization condition}

\author{Tyler D. Blanton\\
        Physics Department, University of Washington, Seattle WA 98195-1560, USA\\
        E-mail: \email{blanton1@uw.edu}}

\author{Ra\'ul A. Brice\~no\\
Department of Physics, Old Dominion University, Norfolk, Virginia 23529, USA\\
\& Thomas Jefferson National Accelerator Facility,
Newport News, VA 23606, USA\\
        E-mail: \email{rbriceno@jlab.org}}
        
\author{Maxwell T. Hanson\\
Theoretical Physics Department, CERN, 1211 Geneva 23, Switzerland\\
E-mail:\email{maxwell.hansen@cern.ch}}

\author{Fernando Romero-L\'opez\\
IFIC, CSIC-Universitat de Val\`encia, 46980 Paterna, Spain
E-mail:\email{fernando.romero@uv.es}}

\author{\speaker{Stephen R. Sharpe}\\
        Physics Department, University of Washington, Seattle WA 98195-1560, USA\\
        E-mail: \email{srsharpe@uw.edu}}

\abstract{We describe recent work on the relativistic three-particle quantization condition,
generalizing and applying the original formalism of 
Hansen and Sharpe, 
and of Brice\~no, Hansen and Sharpe. 
In particular, we sketch three recent developments:
the generalization of the formalism to include K-matrix poles; 
the numerical implementation of the quantization condition in the isotropic 
approximation; 
and ongoing work extending the description of the three-particle divergence-free K matrix beyond the isotropic approximation.}

\FullConference{The 36th Annual International Symposium on Lattice Field Theory - LATTICE2018\\
		22-28 July, 2018\\
		Michigan State University, East Lansing, Michigan, USA.}

\begin{document}

\section{Introduction, motivation, and previous work}
If we are to understand strong-interaction resonances 
from first principles, then it is essential to develop methods that can extract their properties
from the quantities that lattice QCD can calculate, namely the spectrum of states with given 
quantum numbers confined to a (usually cubic) box. While methods for describing resonances
having only two-particle decay channels are well developed, and have been extensively applied in numerical lattice calculations, the formalism for scattering and decays to three or more particles lags behind. The issue is urgent both because most resonances do have such decay channels,
and because numerical simulations are already exploring this regime and need such
formalism to interpret the results.

Examples of resonances 
for which a three-particle formalism is needed are
$\omega\to3 \pi$,  $a_2(1320)\to\rho\pi\to 3\pi$, the Roper resonance
$N(1440)\to \Delta\pi\to N\pi\pi$ and $X(3872)\to J/\Psi \pi\pi$.
As illustrated by these examples, for most resonances
one or more of the two-particle subchannels
is also resonant in the kinematical range of interest. It is this situation which is addressed
in the first of the new results described below.

We also note that a three-particle quantization condition is a step on
the way to using lattice QCD to predict
electroweak decay amplitudes in which the final states either have
three or more particles, e.g. $K\to3 \pi$,
or contain two particles that can mix by the strong interactions with states of
three or more particles,
e.g. $D\to 2\pi$ and $D\to \bar K K$.

Here we discuss the status of our model-independent relativistic approach.
This began in Refs.~\cite{\HSQCa,\HSQCb} with the derivation of the three-particle
quantization condition for identical scalar particles of mass $m$, 
with no resonances in two-particle subchannels,
and satisfying a G-parity-like $Z_2$
symmetry that conserves particle number modulo 2.
The procedure to go from the spectrum of two and three particles in a box to
infinite-volume scattering amplitudes involves two steps:
\begin{itemize}
\item[(i)] 
Use the two-particle spectrum in the energy range $E_{2;\rm CM} < 4 m$
and the three-particle spectrum for $E_{3;\CM} < 5 m$ to 
determine the two-particle K matrix, $\K_2$,  and the three-particle
scattering quantity, $\K_{\df,3}$, using the quantization conditions\footnote{%
The subscript $\df$ stands for ``divergence free". See Ref.~\cite{\HSQCa} for
further explanation.}
\begin{equation}
\det\left[ F_2^{-1} + \K_2\right] = 0\,,\qquad
\det\left[ F_3^{-1} + \K_{\df,3}\right] = 0 \,.
\label{eq:QC1}
\end{equation}
The first of these is the L\"uscher quantization condition~\cite{\Luscher},
or a generalization thereof, in which $F_2$ is a known volume-dependent matrix.
The second quantization condition is the result of Ref.~\cite{\HSQCa},
with $F_3$  a known volume-dependent matrix that depends on $\K_2$.
$\K_{\df,3}$ is an intermediate, regularization-scheme-dependent, infinite-volume
quantity that can be roughly thought of as a quasilocal three-particle interaction.
\item[(ii)] 
Determine the three-particle scattering amplitude $\cM_3$ from $\K_{\df,3}$ and $\K_2$
by solving infinite-volume integral equations~\cite{\HSQCb}.
\end{itemize}
The need for two steps, as opposed to the single-step determination of $\K_2$
in the two-particle case, appears to be general. It holds also for the NREFT approach
of Refs.~\cite{\Akakia,\Akakib}, and the ``finite volume unitarity" approach
of Refs.~\cite{\Mai}, although the details of the second steps differ.

The first major generalization within our approach was to remove the $Z_2$-symmetry
constraint, thus allowing $2\leftrightarrow 3$ transitions.
The quantization condition then becomes~\cite{\BHSQC}
\begin{equation}
\det\left[ \begin{pmatrix} F_2 & 0 \\ 0 & F_3 \end{pmatrix}^{-1}
+ \begin{pmatrix} \K_{22} & \K_{23}\\ \K_{32} & \K_{\df,33} \end{pmatrix} \right]
=0\,.
\label{eq:QC2}
\end{equation}
The finite-volume functions are essentially unchanged, and correspond to on-shell
two- and three-particle intermediate states. The number of infinite-volume intermediate
scattering quantities has enlarged to correspond to the allowed physical scattering
processes. The four $\K$-matrices are again related to the corresponding four physical
amplitudes ($\cM_{22}$, $\cM_{23}$, $\cM_{32}$ and $\cM_{33}$) by 
integral equations. Despite its apparent simplicity, Eq.~(\ref{eq:QC2}) was
challenging to derive due to the need for a careful investigation of $1\to 2$ subprocesses.

\section{Including resonant subchannels}

We have recently taken the final major step in the development of our formalism,
namely the inclusion of resonant subchannels. The results described above assumed
that $\K_2$ had no poles on the real axis, as would be the case, for example,
if there were a nearby complex pole in $\cM_2$ corresponding to a resonance.
We have now removed this restriction, albeit within the context of a $Z_2$-symmetric theory,
and find that the quantization condition becomes~\cite{BHSKpole}
\begin{equation}
\det\left[ \begin{pmatrix} F_{\ttil\ttil}& F_{\ttil 3} \\ F_{3\ttil} & F_{33} \end{pmatrix}^{-1}
+ \begin{pmatrix} \K_{\df,\ttil\ttil} & \K_{\df,\ttil3}\\ \K_{\df,3\ttil} & \K_{\df,33} \end{pmatrix} \right]
=0\,.
\label{eq:QC3}
\end{equation}
This is the result for a single K-matrix pole within the relevant kinematic range,
i.e. $E_{2,\CM} < 4m$. The second channel in this case, labeled $\ttil$, is an unphysical 
resonance $+$ particle channel. It is required in our analysis because the poles in
$\K_2$ lead to power-law finite-volume dependence at intermediate stages, although they
cancel in the final result. All four entries in the $F$ matrix are known in terms of $\K_2$ and
the box size. There is again a second step in which the four intermediate quantities 
(the entries of the $\K_{\df}$ matrix) are related to the single physical scattering
amplitude, $\cM_3$, by integral equations.

There is not space here for further explication of the result (\ref{eq:QC3}), or of its 
derivation. For these, see Ref.~\cite{BHSKpole}. We found this to
be the most challenging of the three results  to derive. We stress, however, that,
as in the previous work,  we do not make assumptions about the form of the effective
field theory that we consider, aside from its $Z_2$ symmetry.

Looking forward, there are several further generalizations on our to-do list: multiple
K-matrix poles, K-matrix poles in a theory without the $Z_2$ symmetry, and allowing for
nondegenerate particles and multiple three-particle channels. In all these cases we are 
confident, based on past experience, that we know the form of the resulting quantization
conditions and that their derivation will be relatively straightforward generalizations of previous
work. Other formal work we are considering is better understanding the relation of our quantization condition to those of the other two approaches, and understanding 
how the results in Eqs.~(\ref{eq:QC2}) and (\ref{eq:QC3}) connect when 
a resonance changes into a stable state, an example being
the $\rho$ meson as the light quark masses increase.

\section{Numerical implementation in the isotropic approximation}

Since the formalism is in a relatively advanced state, the time has come to
learn how to implement it.
Each of the quantities appearing in the quantization conditions is an infinite-dimensional
matrix, and must be truncated for practical applications.
In the two-particle case, this is accomplished by assuming that $\K_2$ vanishes
for partial waves above a chosen cutoff $\ell_{\rm max}$ (usually $0$, $1$ or $2$).
This is appropriate close to threshold, since partial wave amplitudes scale as
$q^{2\ell}$, with $q$ the relative CM momentum.
In addition, the nonvanshing components of $\K_2$ 
must be parametrized in a physically-motivated way, 
and the parameters then determined from fitting the energy levels obtained from
simulations.
This setup has, by now, been widely and successfully implemented for two particles, including
multiple two-particle channels.

We have begun the implementation of a similar truncation scheme for 
$\K_{\df,3}$, applying this first to our original formalism, Eq.~(\ref{eq:QC1}).\footnote{%
We note that this formalism---identical particles with a $Z_2$ symmetry and no subchannel resonances---applies to the case of three pions in an $I=3$ state (assuming exact
isospin symmetry). This three-pion system has recently been analyzed in great detail using
the finite-volume unitarity approach~\cite{Mai:2018djl}. 
See also Ref.~\cite{Romero-Lopez:2018rcb}.
 }
The situation is more complicated than for $\K_2$
because $\K_{\df,3}$ has additional matrix indices: 
a spectator momentum index (a finite-box momentum)
together with the angular momentum indices of the nonspectator pair.
Nevertheless, a systematic expansion about the three-particle threshold is possible,
as discussed in Sec.~\ref{sec:beyond} below.

In Ref.~\cite{\BHSnum}, we have implemented the leading term in this expansion,
first introduced in Ref.~\cite{\HSQCa}, and denoted  the isotropic approximation.
Both $\K_2$ and $\K_{\df,3}$ are truncated at $\ell_{\rm max}=0$,
and $\K_{\df,3}$ is taken to be a point-like vertex, 
independent of the spectator momentum.
Furthermore, only the leading term in an effective range expansion for  $\K_2(\ell=0)$  is kept,
namely the s-wave scattering length, $a_0$.
Thus scattering is described by two parameters, $a_0$ and $\K_{\df,3}$.
Our aim  is not to suggest that this approximation will be sufficient to describe
most resonances (although it may be adequate for the $I=3$ three pion state close to threshold),
but rather to show how the two steps connecting the spectrum to scattering amplitudes
can be implemented from beginning to end.
We note that this truncation is the analog in our formalism of the approximations used
in the other two approaches~\cite{\Akakia,\Akakib,\Mai,Doring:2018xxx,Mai:2018djl}.

In the isotropic approximation, the quantization condition of Eq.~(\ref{eq:QC1}) collapses
to a one-dimensional algebraic equation, although it depends on all components of
$F_3$ up to a cutoff $k\sim m$. It turns out to be straightforward to implement numerically for
a wide range of box sizes $4 \lesssim mL \lesssim 70$.
The motivation for considering extremely large volumes is twofold:
first, it allows us to provide robust checks of asymptotic predictions;
and, second, it provides a tool for solving the integral equations relating $\K_{\df,3}$ to $\cM_3$,
although so far only below or at the three-particle threshold.

We have space here only to mention a few of the many results presented
in Ref.~\cite{\BHSnum}.
Our most straightforward investigation is to determine how the three-particle spectrum 
depends on $a_0$ and $\K_{\df,3}$. The noninteracting
energy levels are shifted 
in the expected way: attractive two- and three-particle interactions lower the energies,
while repulsive interactions raise them. For small enough $|a_0 m|$ and $(mL)^{-1}$,
we can check our methods by comparing to the threshold expansion derived analytically in Ref.~\cite{\HSTH}. In this regime the contribution of $\K_{\df,3}$ is suppressed by
$(mL)^{-3}$ compared to the leading term.
Our approach works, however, for arbitrarily large negative $a_0 m$, i.e. with a strongly
attractive interaction that does not quite lead to a two-particle bound state.
In this regime we find that the shift from noninteracting energy levels is comparable to the
level spacing.
Furthermore, we find that if $mL\sim 5$, which is the regime accessed by
present numerical simulations, the energy levels depend significantly on $\K_{\df,3}$.
This suggests that it will be possible to extract information on the three-particle
interaction, despite its suppression for large volumes.

A nice example of the full two-step application of our formalism concerns the properties
of an Efimov-like three-particle bound state. We expect such bound states near the
unitary limit $a_0 m\to -\infty$, and find that for each choice of $|a_0 m| \gg 1$, there
is a corresponding value of $\K_{\df,3}$ for which there is a bound state.
We choose $a_0 m= - 10^{4}$ for detailed investigation, finding the bound-state energy
$E_B=2.98858 m$.
For this system, Ref.~\cite{\AkakiBS} has derived the asymptotic volume dependence
analytically, and we find that our results match this prediction, 
although  the asymptotic regime begins only at $mL\sim 40$, 
well beyond the reach of simulations. 
For acheivable values of $mL$, the quantization condition predicts very large
energy shifts, e.g. $E\approx 2.65 m$ at $mL=5$.
This is thus an example of the utility of the quantization condition---one can use it
to extract the underlying scattering parameters in a regime where analytic results
are not available.

\begin{figure}[tbh]
\begin{center}
\includegraphics[width=0.8\textwidth]{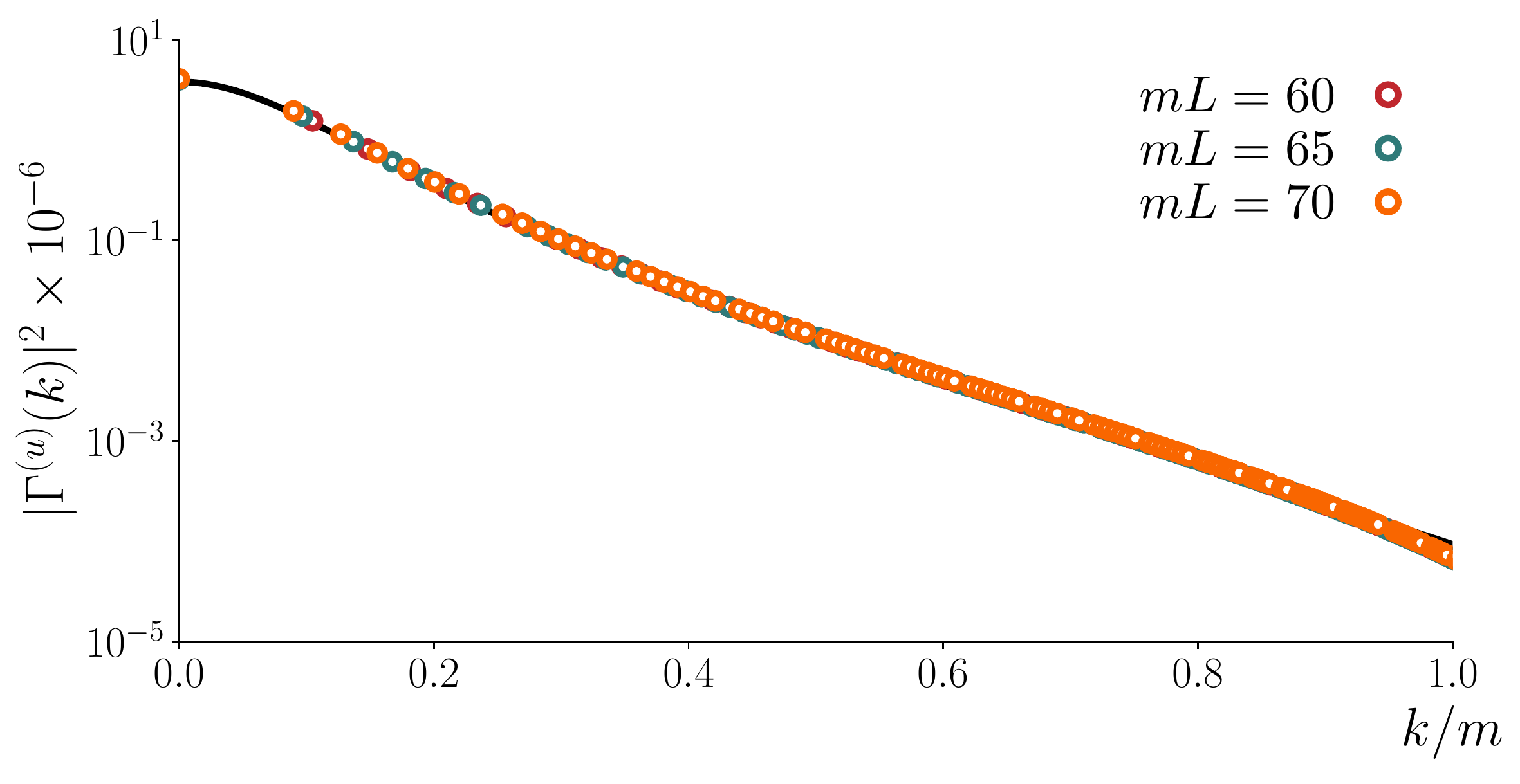}
\caption{
Momentum dependence of the {magnitude squared of the} bound-state residue function. 
The curve shows the theoretical prediction, while the points are the results from 
solving the integral equations. Here $1/(mL)$ is proportional to the grid spacing 
used in the discretization of
the integral equation, and the agreement between the results for $mL=60-70$ shows
that the dependence on this spacing has been removed.
\label{fig:BS} }
\end{center}
\end{figure}

Implementing the second step in our approach allows us to study $\cM_3$ close
to the bound state, where it has a pole,
and to determine the dependence of the residue on the spectator momentum $k$.
This is related to the momentum-space wavefunction of the state.
In Ref.~\cite{\HSBS} we have extended the work of Ref.~\cite{\AkakiBS}
to determine the analytic form for this quantity using NRQM. 
Given the fit to the asymptotic behavior of $E_B(L)$, we then have a theoretical
{\em parameter-free} prediction for the residue function. As shown in Fig.~\ref{fig:BS},
our numerical results for this quantity agree with the prediction over many orders
of magnitude. This demonstrates that our formalism can reproduce known, nontrivial
physics, and also shows how, in principle, one can go from simulations to physical predictions.

\section{Beyond the isotropic approximation}
\label{sec:beyond}

We close by describing briefly work in progress that develops a systematic expansion 
for $\K_{\df,3}$ about threshold, and shows how this can be implemented in practice~\cite{BRSinprogress}.
We continue to work in the $Z_2$ symmetric theory without K-matrix poles.
What we aim for is an expansion for $\K_{\df,3}$ that is the analog of
the threshold (or effective range) expansion for $\K_2$. The strategy was described
in Ref.~\cite{\BHSnum}. The key inputs are that $\K_{\df,3}$ is a relativistically invariant
function of the three incoming and three outgoing momenta ($p_i$ and $p'_i$, 
respectively, with $i=1,3$),\footnote{%
In the formalism described in Refs.~\cite{\HSQCa,\HSQCb}, $\K_{\df,3}$ was not
relativistically invariant. However, as was noted in Ref.~\cite{\BHSQC}, by a small
change in the definition of $F_3$, it can be made so.}
and is symmetric separately 
under interchange of the initial the final momenta.

Using these results we can expand $\K_{\df,3}$ about threshold. The relevant
variables are
\begin{equation}
\Delta=s-9m^2\,,\ \
\Delta_i = (p_j+p_k)^2-4m^2\,,\ \
\Delta'_i = (p'_j+p'_k)^2-4m^2\,,\  {\rm and}\ 
t_{ij} = (p_i-p'_j)^2\,,
\label{eq:thrquant}
\end{equation}
where $s=(p_1+p_2+p_3)^2$, and, in the second and third equations,
$i,j,k$ are one of the three cyclic permutations of $1,2,3$.
All  16 quantities in Eq.~(\ref{eq:thrquant})
vanish at threshold and we treat them as being of the same order in the expansion.
Only eight of them are independent, corresponding to the
overall energy and the seven angular variables that describe $3\to3$ scattering.
Implementing the symmetries, we find that the expansion to quadratic order in small 
quantities is~\cite{BRSinprogress}
\begin{align}
\K_{\df,3} &= c_0 + c_1\Delta + c_2\Delta^2 + c_A\K_{2A} + c_B\K_{2B}\,,
\\
\K_{2A} &= \sum_{i=1}^3 \left(\Delta_i^2 + {\Delta'_i}^2\right)\,,
\quad
\K_{2B} = \sum_{i,j=1}^3 t_{ij}^2\,.
\end{align}
At leading order, there is a single constant, $c_0$---this we 
called $\K_{\df,3}$ in the previous section.
At linear order there is also a single constant, $c_1$, which leads to linear dependence
on $s$, but no angular dependence.
What is new and striking is that, at quadratic order, there are, in addition to the $s^2$ term,
only two terms that depend on angles, $\K_{2A}$ and $\K_{2B}$.
There are thus five relations between the angles at this order,
a simplification that makes implementation significantly more tractable.

We have converted the expressions $\K_{2A}$ and $\K_{2B}$ into the variables used
in the quantization condition: the spectator momentum $k$ and angular momentum
for the remaining pair, $\ell,m$. We find contributions involving both $\ell=0$ and $2$.
For consistency, this means that we must include d-wave two-particle
K-matrices in addition to s-wave.\footnote{%
Odd angular momenta vanish for identical particles.}
At quadratic order, these take the form
\begin{equation}
\frac1{\K_2^s} =\frac1{16\pi E_{2}}\left[\frac1{a_0} +r_0 \frac{q^2}{2}
+ P_0 r_0^3 q^4 \right]\,,
\qquad
\K_2^d={16\pi E_{2}} q^4 a_2^5 \,,
\end{equation}
where $E_2$ is the two-particle CM energy and $r_0$ is the effective range.
We have parametrized $\K_2^d$ so that $a_2$ has dimensions of length. 
Thus, altogether there are nine parameters at quadratic order.\footnote{%
In Ref.~\cite{\HSQCa} we defined the isotropic approximation to mean that 
$\K_{\df,3}$ can have arbitrary dependence on $s$, but no dependence on angles.
We now see that, beyond linear order in $s$,
one should introduce angular dependence as well
in order to be part of a systematic threshold expansion.}

We have numerically implemented the extended quantization condition, 
including projection onto irreps of the cubic group.
Here all irreps appear, whereas in the isotropic approximation only the trivial irrep, $A_1^+$,
contributes.
We are presently investigating its properties in detail, and hope to report results soon.

\section{Acknowledgments}
The work of TDB and SRS was supported in part by the United States Department of Energy grant No.~DE-SC0011637. RAB acknowledges support from U.S. Department of Energy contract DE-AC05-06OR23177, under which Jefferson Science Associates, LLC, manages and operates Jefferson Lab.  FRL acknowledges the support of the European Project InvisiblesPlus H2020-MSCA-RISE-2015. The work of FRL has also received funding from the European Union Horizon 2020 research and innovation program under the 
Marie Sklodowska-Curie grant agreement No. 713673.

\bibliographystyle{apsrev4-1}
\bibliography{ref}


\end{document}